\documentstyle{article}
\input{epsf}
\def\dofig#1#2{\epsfysize=#1 \centerline{\epsfbox{#2}}}

\begin{document}

\title{Oscillatory correlation of delayed random walks}
\author{Toru Ohira\\
 Sony Computer Science Laboratory\\ 
3-14-13 Higashi-gotanda,\\
Tokyo 141, Japan\\
ohira@csl.sony.co.jp\\  
\\
\\
Sony Computer Science Lab Technical Report: SCSL-TR-96-014
}

\date{\ }

\maketitle
\thispagestyle{empty}

\begin{abstract}
We investigate analytically and numerically
the statistical properties of a random walk model with delayed transition
probability dependence (delayed random walk).
The characteristic feature of such a model is the oscillatory
behavior of its
correlation function. 
We investigate a model
whose transient and stationary oscillatory behavior is 
analytically tractable.
The correspondence of the model with a Langevin equation with
delay is also considered.
\end{abstract}

Noise and correlative effects (memory) are two elements
which are associated with many natural systems. 
In physics, two main approaches have been 
developed to study such systems
with noise and memory.
One approach is formulating the model in physical
space with a differential equation of motion such as the
``generalized Langevin equation''\cite{kubo,mori}. 
The other 
is to formulate a model in probability space
as a non-Markovian problem as in the ``generalized master
equation'' approach \cite{gmaster}.
 These two avenues have been developed and applied to various problems 
in physics. Examples include studies on the Alder-Wainwright 
effect\cite{okabe},
spin relaxation\cite{desposito}, and driven two-level atoms\cite{brinati}.
 
The delayed stochastic system we discuss here can be viewed as a special case,
where only a single (memory) point at a fixed time interval
in the past has
influence on the current state of the system. 
Research of such systems, particularly those with no noise, has
been carried out in fields of 
mathematics\cite{cooke}, biology\cite{biodelay},
artificial neural network\cite{marcus},
electrical circuits\cite{losson},
as well as in physics\cite{chaoscont}.
Models with both noise and delay have also
been considered numerically\cite{bionoisedelay} and analytically as an extention
of the Langevin equation\cite{kuchler}. 
These works represent approaches and formulations
in physical space. For the probability space approach,
 ``Delayed random
walk'' is recently proposed \cite{ohira}
 and has been applied to model human posture controls\cite{collins}. 
However, an analytical understanding of this random walk
is yet far from being complete.   
 
The main theme of this paper is to increase the analytical understanding
of the behavior of a delayed random walk
model. The oscillatory correlation
function is found to be associated with delayed random walks \cite{ohira,ohira2}. 
We show here that such oscillatory behavior of the of the
correlation function is analytically tractable.
From the study of random walks point of view,
this delayed random walk model provides an example
whose correlation function behaves differently
compared to commonly known random walks with memory,
such as self-avoiding, or persistent walks \cite{weiss}. 
In addition,
we note that 
oscillatory or chaotic behavior associated with
delays are generally 
difficult to analyze\cite{bionoisedelay}. 
Hence, this model
also serves as one of the rare analytically tractable examples among  
models with delay.

We consider a random walk which takes a unit step in a unit time.
The delayed random walk we start with is an extension of a position 
dependent random walk whose step toward the origin is more likely
when no delay exists. 
Formally, it has the following definition:
\begin{eqnarray}
P(X_{t+1}&=&n; X_{t+1-\tau}=s)\nonumber\\
 &=& g(s-1)P(X_{t}=n-1;X_{t+1-\tau}=s;X_{t-\tau}=s-1)\nonumber\\ 
 &+& g(s+1)P(X_{t}=n-1;X_{t+1-\tau}=s;X_{t-\tau}=s+1)\nonumber\\
 &+& f(s-1)P(X_{t}=n+1;X_{t+1-\tau}=s;X_{t-\tau}=s-1)\nonumber\\
 &+& f(s+1)P(X_{t}=n+1;X_{t+1-\tau}=s;X_{t-\tau}=s+1),
\label{drw}\\
f(x) & + & g(x) = 1,
\end{eqnarray}
where the position of 
the walker at time $t$ is $X_{t}$, and $P(X_{t_1}=u_1;X_{t_2}=u_2)$ 
is the joint probability
for the walker to be at $u_1$ and $u_2$ at time $t_1$ and $t_2$, respectively. 
$f(x)$ and $g(x)$ are transition probabilities to take a step to the negative and
positive directions respectively at the position $x$. 
In this paper, we further place the conditions:
\begin{equation}
f(x)  > g(x) \quad (x>0),\quad \quad
f(-x) = g(x) \quad (\forall x).
\end{equation}
These conditions make the delayed random walks symmetric with 
respect to the origin, which is attractive without delay $(\tau=0)$.

We now proceed to obtain a few 
properties from this general definition.
By the symmetry with respect to the origin,
the average position of the walker is $0$.
This symmetry is further 
used to inductively show \cite{ohira3} in the stationary state
($t \rightarrow  \infty$) 
that
\begin{equation}
P(X_{t+1}=n;X_{t}=n+1) = P(X_{t+1}=n+1;X_{t}=n). 
\end{equation}
We derived the stationary probability distribution
for the previously discussed delayed random walk
model using this property \cite{ohira}.
Also, the multiplication of Eq. (\ref{drw}) for the stationary state
 by $\cos(\alpha n)$ and summation over
$n$ and $s$ yields for the generating function:
\begin{equation}
\langle \cos(\alpha X_{t}) \rangle =  
\cos(\alpha)\langle \cos(\alpha X_{t}) \rangle 
+ \sin(\alpha)\langle \sin(\alpha X_{t}) 
\{ f(X_{t-\tau})-g(X_{t-\tau}) \} \rangle
\end{equation}
In particular, we have a following invariant
relationship with respect to the delay. 
\begin{equation}
{1 \over 2} = \langle X_{t}\{ f(X_{t-\tau})-g(X_{t-\tau}) \} \rangle
\label{inv}
\end{equation}
This invariant property is used below. 

We will consider a specialized model for the rest of this
paper \cite{note1}. We define $f(x)$ and $g(x)$ as
\begin{eqnarray}
f(x) &= &{1 \over 2}(1 + 2d) \quad (x>a),\quad\quad 
      {1 \over 2}(1 + \beta x)\quad (-a \leq x \leq a),\quad\quad{1 \over 2}(1 - 2d) \quad (x<-a),\nonumber\\
g(x) &= &{1 \over 2}(1 - 2d) \quad (x>a),\quad\quad 
     {1 \over 2}(1 - \beta x)\quad (-a \leq x \leq a),\quad\quad{1 \over 2}(1 + 2d) \quad (x<-a).\nonumber\\
\end{eqnarray}
Physically, this model implies that when $\tau = 0$ the transition 
probability for the walker to move toward the origin 
increases
linearly at a rate of $\beta \equiv d/a$ as the 
distance increases from the
origin
up to the potition $a$ after which the transition 
probability is held constant.
We assume that with sufficiently large $a$, we can
ignore the probability for the walker to be outside of the
range (-a, a). 

Then, the previous invariant 
relation in Eq. (\ref{inv}) becomes the following with this model:
\begin{equation}
\langle X_{t}X_{t-\tau} \rangle  = K(\tau) = {1 \over {2\beta}}.
\label{taucorr}
\end{equation}
This invariance with respect to $\tau$ of
the correlation function with $\tau$ steps apart is a simple characteristic
of this delayed random walk model. This property is a key to obtaining the
analytical expression for the correlation function, which we
now turn our attention.

For the stationary state and $0 \leq u \leq \tau$, the following
is obtained from the definition (\ref{drw}).
\begin{eqnarray}
P(X_{t}=n; X_{t-u}=l) &=& \sum_{s} g(s)P(X_{t}=n-1;X_{t-(u-1)}=l;X_{t-\tau}=s) 
\nonumber \\
&+& \sum_{s} f(s)P(X_{t}=n+1;X_{t-(u-1)}=l;X_{t-\tau}=s)
\end{eqnarray}
We can derive the following equation for the correlation function by multiplication
of this equation by $nl$ and summing over.
\begin{equation}
K(u) = K(u-1) - \beta K(\tau+1-u), \quad\quad ( 0 \leq u \leq \tau ).
\label{corr1}
\end{equation}
A similar argument gives for $\tau < u$,
\begin{equation}
K(u) = K(u-1) - \beta K(u -1 - \tau), \quad\quad ( \tau < u ).
\label{corr2}
\end{equation}
Equations (\ref{corr1}) and (\ref{corr2}) can be solved explicitly
using (\ref{taucorr}). In particular, for 
$0 \leq u \leq \tau$  we obtain
\begin{eqnarray}
K(u) &=& K(0){ {( m_{+}^{u} - m_{+}^{u-1} ) 
- ( m_{-}^{u} - m_{-}^{u-1} )}\over{
 m_{+} - m_{-} } } - {1 \over 2}{ {( m_{+}^{u} 
- m_{-}^{u} )}\over{ m_{+} - m_{-
} } }\nonumber\\
K(0) &=& {1 \over {2 \beta}} { { (m_{+} - m_{-}) + \beta ( m_{+}^{\tau} - m_{-}^
{\tau} )} \over {( m_{+}^{\tau} - m_{+}^{\tau-1} ) 
- ( m_{-}^{\tau} - m_{-}^{\tau
-1} )} } \nonumber\\
m_{\pm} &=&  {(1 - {\beta^2 \over 2})} \pm {\beta \over 2}\sqrt{ \beta^2 - 4 }  
\label{drwsol}
\end{eqnarray}
For $\tau < u$, it is possible to write $K(u)$ in a multiple summation
form, though the expression becomes rather complex. For example,
with $\tau < u \leq 2\tau$,
\begin{equation}
K(u) = {1\over 2\beta} - \beta \sum_{i=0}^{u-1-\tau}K(i) \nonumber
\end{equation}
where $K(i)$ summed is given by the equation (\ref{drwsol}).
\clearpage

The behavior of the correlation function is shown in
Fig. 1. As we increase $\tau$, oscillatory behavior
of the correlation function appears. 
The decay of the peak envelope is found
numerically to be exponential. The decay rate of the
envelope for the
small $u$ is approximately $1/(2K(0))$. Also we note the mean
square postion ($K(0)$) increases with increasing delay
$\tau$.
\begin{figure}[h]
\dofig{3.0in}{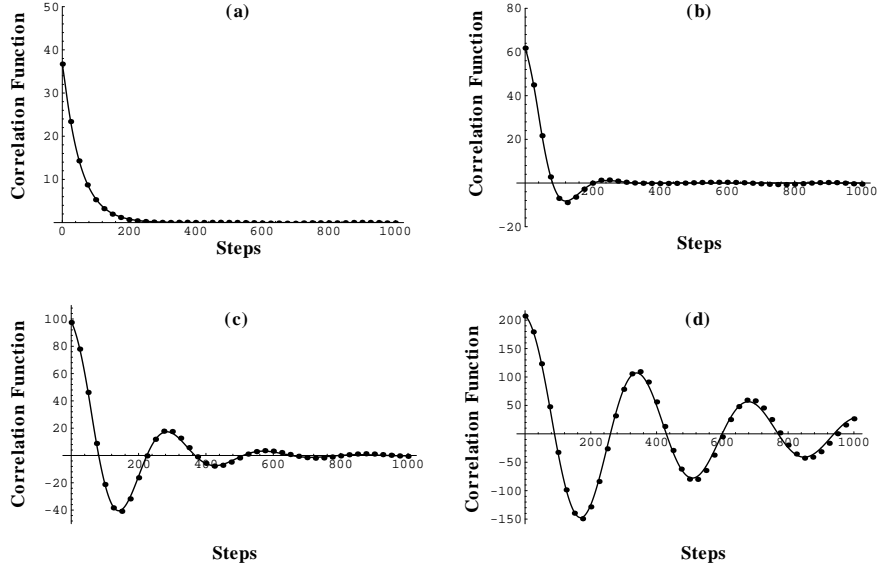}
\caption{ Stationary correlation function $K(u)$ from 
simulations (dots) as a function of steps $u$ with varying 
$\tau$ compared with the analytical solution 
obtained in the text (line). 
The parameters are set as $a=50$, $d = 0.4$,
and
$\tau = (a) 10, (b) 40, (c) 60, (d) 80$.
The simulation performed random walks of 6000 steps
starting from the origin. 
The position data after 4500 steps are used
to compute the correlation and averaged over 10,000 trials.}
\end{figure}

Analysis of the correlation function for the transient state
can be done in the similar argument as in the stationary state.
We can derive the set of coupled dynamical equations as 
follows:
\begin{eqnarray}
K(0,t+1) & = & K(0,t) + 1 - 2\beta K(\tau,t)\nonumber\\
K(u,t+1) & = & K(u-1,t) - \beta K(\tau-(u-1),t+1-u),\quad (1\leq u \leq \tau)\nonumber\\ 
K(u,t+1) & = & K(u-1,t) - \beta K((u-1)-\tau,t+1-u),\quad (u > \tau) 
\label{dynaeq}
\end{eqnarray}
For the initial condition, we need to specify the correlation function for the
interval of initial $\tau$ steps. Let us consider a random walk, which is held
at the origin before it begins to take a steps, thus performing a homogeneous
random walk for the steps $(1, \tau)$. This translates to the initial condition
for the correlation function as
\begin{equation}
K(u, t) = t - u \quad (0 \leq u \leq \tau).
\end{equation}
The solution can be iteratively generated for Eq. (\ref{dynaeq}) given
this initial condition. We have plotted some examples for the dynamics of the
mean square displacement $K(0)$ in Figure 2.
Again, the oscillatory behavior arises with increasing $\tau$.
Hence, in the model discussed here 
the oscillatory behavior with increasing
delay appears in both its stationary and transient states.
\begin{figure}
\dofig{2.2in}{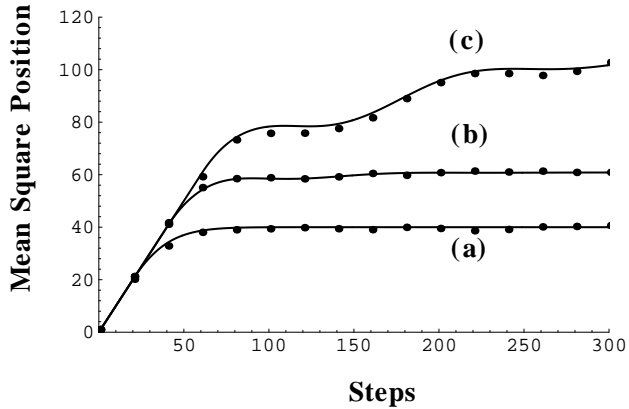}
\caption{ Examples of dynamics of the mean square position
${\langle X^2 \rangle} = K(0)$  
with varying 
delay $\tau$.
The data is from simulations (dots) averaged over 10,000 trials,
and from the analytical solutions (line).
The parameters are set as $a=50$, $d=0.45$, and 
$\tau = (a) 20, (b) 40, (c) 60$.}
\end{figure}

Let us now briefly discuss relationship of this model to 
the Langevin equation with delay:
\begin{equation}
{d \over dt}X_{t} = - \beta X_{t-\tau} + \xi_{t}, \quad
\langle \xi_{t_1}\xi_{t_2}\rangle = \delta(t_1 - t_2).
\label{dle}
\end{equation}
This Langevin equation is a special case of
the equation considered in \cite{kuchler}. 
It should be noted that the equation is
normalized with the ``width'' of the noise $\xi_{t}$.
It has been shown that:

(1) the equation is stationary if and only if $\tau< {\pi \over {2 \beta}}$.

(2) The stationary
 correlation function $K(r) \equiv \langle X_{t}X_{t-r}\rangle$ 
has the following form when $r < \tau$:
\begin{equation}
K(r)= K(0)\cos(\beta r) - {1 \over 2 \beta} \sin(\beta r), \quad
K(0)=  { {1 + \sin(\beta \tau)} \over {2 \beta \cos(\beta \tau)} }
\label{km93}
\end{equation}

When $\beta<<1$ (or $a>>d$), the delayed random walk
model approximately corresponds to this Langevin equation with
delay.
In particular, we can obtain Eq. (\ref{km93}) from
the result (\ref{drwsol}) obtained for the delayed random walk,
by expanding in small $\beta$.

Some points of discussion are now in order. 
The first point is how this model is placed
in relation to other models with noise and
delay (or memory, to be more general).
In particular we 
 note that the Langevin equation discussed here is
not a special case of the generalized Langevin equation
which is consistent with the fluctuation-dissipation theorem.
As argued in \cite{kubo} for the generalized
Langevin equation, the noise term needs
to be ``colored'' in equation (\ref{dle}) for
the consistency. Investigation of the colored noise
case in its relation to delayed random walks as
well as further studies of the correspondence of 
dynamical aspects of (\ref{drw})
and (\ref{dle}) are 
currently underway.  
Finally, we ask what the possible applications are of delayed 
random walks. As mentioned before model with a different transition property 
has been applied \cite{ohira} to describe the qualitative statistical
behavior of the center of gravity in a human posture control
experiment \cite{collins}. Also, oscillatory correlation functions appear
in such numerical studies of phase separation dynamics under
stirring \cite{lacasta} and of response dynamics of
neural recepter cell syncytium \cite{funakubo}.  
Applications or relations to these and other systems are
currently being sought and considered.

The model provided here is simple, yet has shown
some characteristics of systems with noise and delay.  
It is hoped that further investigation of this and
extended delayed random walks will provide us deeper understanding of 
delayed stochastic systems.

The author would like to thank Drs. Y. Okabe and K. Umeno
for discussions, and J. Milton and M. Mackey 
for suggestions of references.


\begin{thebibliography}{99}
\bibitem{note1}
This restriction for $f(x)$ and $g(x)$ to be lineary dependent on
$x$ is necessary for the following analytical treatment
for the correlation function.
\end{thebibliography}
\end{document}